\documentclass[twocolumn,showpacs,preprintnumbers,amsmath,amssymb,pre]{revtex4}

\usepackage{graphicx}
\usepackage{dcolumn}
\usepackage{bm}
\usepackage{enumerate}
\usepackage{amsmath}
\usepackage{color}

\begin{document}

\title{$G$-subdiffusion equation that describes transient subdiffusion}

\author{Tadeusz Koszto{\l}owicz}
 \email{tadeusz.kosztolowicz@ujk.edu.pl}
 \affiliation{Institute of Physics, Jan Kochanowski University,\\
         Uniwersytecka 7, 25-406 Kielce, Poland}

\author{Aldona Dutkiewicz}
 \email{szukala@amu.edu.pl}
 \affiliation{Faculty of Mathematics and Computer Science,\\
Adam Mickiewicz University, Uniwersytetu Pozna\'nskiego 4, 61-614 Pozna\'n, Poland}

\date{\today}

\begin{abstract}
A $g$--subdiffusion equation with fractional Caputo time derivative with respect to another function $g$ is used to describe a process of a continuous transition from subdiffusion with parameters $\alpha$ and $D_\alpha$ to subdiffusion with parameters $\beta$ and $D_\beta$. The parameters are defined by the time evolution of the mean square displacement of diffusing particle $\sigma^2(t)=2D_i t^i/\Gamma(1+i)$, $i=\alpha,\beta$. The function $g$ controls the process at ``intermediate'' times. The $g$--subdiffusion equation is more general than the "ordinary" fractional subdiffusion equation with constant parameters, it has potentially wide application in modelling diffusion processes with changing parameters. 
\end{abstract}

\maketitle

Subdiffusion occurs in media, such as gels and bacterial biofilm, where the movement of molecules is very hindered due to a complex structure of a medium \cite{bg,mk,mk1,ks,hughes,jeon,godec,cano,cherstvy,lieleg,wong,km,kmwa,kdm}. Within the Continuous Time Random Walk (CTRW) model subdiffusion is defined as a process in which a time distribution between particle jumps $\psi$ has a heavy tail which makes the average time infinite, $\psi(t)\sim 1/t^{1+\alpha}$ when $t\rightarrow\infty$, $0<\alpha<1$, and the jump length distribution has finite moments \cite{mk,mk1,ks,hughes,mks,skb,sk,barkai2000,compte}, the citation list on the above issues can be significantly extended. This model shows that subdiffusion with a constant subdiffusion parameter (exponent) $\alpha$ in a one--dimensional homogeneous system can be described by an ``ordinary'' subdiffusion equation with a fractional time derivative of the order $\alpha\in(0,1)$
\begin{equation}\label{eqI1}
\frac{^C \partial^{\alpha} C(x,t)}{\partial t^\alpha}=D_\alpha\frac{\partial^2 C(x,t)}{\partial x^2},
\end{equation}
where the Caputo fractional derivative is defined here as 
\begin{equation}\label{eqI2}
\frac{^Cd^{\alpha} f(t)}{dt^\alpha}=\frac{1}{\Gamma(1-\alpha)}\int_0^t (t-u)^{-\alpha}f'(u)du,
\end{equation}
$D_\alpha$ is a generalized diffusion coefficient measured in the units of ${\rm m^2/s^\alpha}$, $C$ is a concentration of diffusing particles, $f'$ denotes the first--order derivative of function $f$. Eq. (\ref{eqI1}) can be transformed to its equivalent form with the fractional Riemann--Liouville time derivative of the order $1-\alpha$, see for example \cite{mk}. 

Subdiffusion parameters are often defined by a time evolution of the mean square displacement $\sigma^2$ of a diffusing particle, 
\begin{equation}\label{eqI3}
\sigma^2(t)=\frac{2D_\alpha t^\alpha}{\Gamma(1+\alpha)}.
\end{equation} 
Eq. (\ref{eqI1}) describes the subdiffusion process with constant parameters $\alpha$ and $D_\alpha$. Such a process can occur in a homogeneous system in which the structure does not change with time. However, the structure of a medium may evolve over time continuously changing the parameters. An example is antibiotic subdiffusion in a bacterial biofilm \cite{km,kmwa}. Bacteria have different defense mechanisms against the action of the antibiotic, which can slow down or even significantly accelerate the antibiotic transport \cite{aot,mot}. Different models have been used to describe subdiffusion with variable parameters \cite{hanes,roth,chen2013}. Subdiffusion equations with linear combination of fractional time derivative with different parameters $\alpha$ have been studied in \cite{sandev2018,sandev,awad}. The transmogrifying CTRW model describing anomalous diffusion with changing subdiffusion parameters has been considered in \cite{angstmann}. Modification of a time scale in a diffusion model can lead to changes in diffusion parameters as well as in the type of diffusion \cite{hilferanton,kd2021a}. A timescale changing can be made by means of subordinated method \cite{ks,sokolov,feller,csm,dybiec}. Within this method retarding and accelerating anomalous diffusions have been obtained \cite{stanislavsky2020,stanislavsky2019}. Examples of processes that lead to a rescaling of diffusion are diffusing diffusivities where the diffusion coefficient evolves over time \cite{csm}, passages through the layered media \cite{carr}, and anomalous diffusion in an expanding medium \cite{levot}. We mention that distributed order of fractional derivative in subdiffusion equation can lead to delayed or accelerated subdiffusion \cite{chechkin2002,chechkin2008,orzel,eab,eab1}. 

We consider subdiffusion in a one--dimensional homogeneous system, diffusive properties of a medium may change over time. At the initial moment the subdiffusion parameters are $\alpha$ and $D_\alpha$, and after long time (formally $t\rightarrow\infty$) the parameters are $\beta$ and $D_\beta$. In these cases subdiffusion is described by the "ordinary" subdiffusion equation. In the ``intermediate'' time interval there is a continuous transient subdiffusion process in which the subdiffusion parameters are not defined by Eq. (\ref{eqI3}). We call the process transient subdiffusion, it is symbolically written as $(\alpha,D_\alpha)\rightarrow(\beta,D_\beta)$.

Recently, the $g$--subdiffusion process characterized by parameters $\alpha$, $D_\alpha$, and by the function $g$ has been considered in \cite{kd2021a,kd2021b}. This process is related to "ordinary" subdiffusion with the same parameters in which the time variable has been rescaled by a deterministic function $g$ which fulfils the conditions $g(0)=0$, $g(\infty)=\infty$, and $g'(t)>0$ for $t>0$, the values of the function $g$ are given in a time unit. $G$--subdiffusion is described by the following $g$--subdiffusion equation 
\begin{equation}\label{eqII1}
\frac{^C \partial^{\alpha}_g C(x,t)}{\partial t^\alpha}=D_\alpha\frac{\partial^2 C(x,t)}{\partial x^2},
\end{equation}
where 
\begin{equation}\label{eqII2}
\frac{^Cd^{\alpha}_g f(t)}{dt^\alpha}=\frac{1}{\Gamma(1-\alpha)}\int_0^t (g(t)-g(u))^{-\alpha}f'(u)du
\end{equation}
is the $g$-Caputo fractional derivative of the order $\alpha\in(0,1)$ with respect to the function $g$ \cite{almeida}. When $g(t)\equiv t$, the $g$-Caputo fractional derivative takes the form of the ``ordinary'' Caputo derivative. We will show that transient subdiffusion can be treated as a special case of $g$--subdiffusion.  We mention that the solutions to the $g$--subdiffusion equation well describe the experimental results of drug diffusion in a system containing tightly packed beads impregnated with the drug at the initial moment \cite{kdlwa}.

The $g$--subdiffusion equation can be solved by means of the $g$--Laplace transform method. The $g$-Laplace transform is defined as \cite{jarad1}
\begin{equation}\label{eqII3}
\mathcal{L}_g[f(t)](s)=\int_0^\infty {\rm e}^{-s g(t)}f(t)g'(t)dt.
\end{equation}
The $g$--Laplace transform is related to the ``ordinary'' Laplace transform $\mathcal{L}[f(t)](s)=\int_0^\infty{\rm e}^{-st}f(t)dt$ as follows
\begin{equation}\label{eqII4}
\mathcal{L}_g[f(t)](s)=\mathcal{L}[f(g^{-1}(t))](s).
\end{equation}
Eq. (\ref{eqII4}) provides the rule
\begin{equation}\label{eqII5}
\mathcal{L}_g[f(t)](s)=\mathcal{L}[h(t)](s)\Leftrightarrow f(t)=h(g(t)).
\end{equation}
The above formula is helpful in calculating the inverse $g$--Laplace transform if the inverse "ordinary" Laplace transform is known. The examples of inverse $g$--Laplace transforms are \cite{kd2021b}
\begin{equation}\label{eqII6}
\mathcal{L}_g^{-1}\left[\frac{1}{s^{1+\nu}}\right](t)=\frac{g^\nu(t)}{\Gamma(1+\nu)}\;,\;\nu>-1,
\end{equation}
\begin{eqnarray}\label{eqII7}
\mathcal{L}_g^{-1}[s^\nu {\rm e}^{-as^\mu}](t)\equiv f_{\nu,\mu}(g(t);a)\\
=\frac{1}{g^{1+\nu}(t)}\sum_{k=0}^\infty \frac{1}{k!\Gamma(-\nu-\mu k)}\left(-\frac{a}{g^\mu(t)}\right)^k,\nonumber
\end{eqnarray}
$a,\mu>0$. The function $f_{\nu,\mu}$ is a special case of the Wright function and the H-Fox function.

The calculations for solving Eq. (\ref{eqII1}) by means of the $g$--Laplace transform method are similar to those for solving Eq. (\ref{eqI1}) using the "ordinary" Laplace transform.
Due to the relation \cite{jarad1}
\begin{equation}\label{eqII8}
\mathcal{L}_g\left[\frac{^C d^\alpha_g f(t)}{dt^\alpha}\right](s)=s^\alpha \mathcal{L}_g[f(t)](s)-s^{\alpha-1}f(0),
\end{equation}
where $0<\alpha\leq 1$, the $g$--Laplace transform of Eq. (\ref{eqII1}) reads
\begin{eqnarray}\label{eqII9}
s^\alpha \mathcal{L}_g[C(x,t)](s)-s^{\alpha-1}C(x,0)\\
=D_\alpha\frac{\partial^2 \mathcal{L}_g[C(x,t)](s)}{\partial x^2}.\nonumber
\end{eqnarray}

The Green's function $P(x,t|x_0)$ is interpreted as a probability density of finding a diffusing particle, located initially at $x_0$, at point $x$ at time $t$. 
The $g$--Laplace transform of Green's function is the following solution to Eq. (\ref{eqII9}) for the initial condition $P(x,0|x_0)=\delta(x-x_0)$, where $\delta$ denotes the delta--Dirac function, and the boundary conditions $\mathcal{L}_g[P(\pm\infty,t|x_0)](s)=0$,
\begin{equation}\label{eqII10}
\mathcal{L}_g[P(x,t|x_0)](s)=\frac{1}{2\sqrt{D_\alpha}s^{1-\alpha/2}}\;{\rm e}^{-\frac{|x-x_0|}{\sqrt{D_\alpha}}s^{\alpha/2}}.
\end{equation}
From Eqs. (\ref{eqII7}) and (\ref{eqII10}) we obtain
\begin{equation}\label{eqII11}
P(x,t|x_0)=\frac{1}{2\sqrt{D_\alpha}}f_{-1+\alpha/2,\alpha/2}\left(g(t);\frac{|x-x_0|}{\sqrt{D_\alpha}}\right).
\end{equation}
Eqs. (\ref{eqII6}) and (\ref{eqII10}) provide 
\begin{equation}\label{eqII12}
\sigma^2(t)=\frac{2D_\alpha}{\Gamma(1+\alpha)}g^\alpha(t).
\end{equation}
Putting $g(t)\equiv t$ in Eq. (\ref{eqII11}) we get the Green's function for the ``ordinary'' subdiffusion equation
\begin{equation}\label{eqII13}
P(x,t|x_0)=\frac{1}{2\sqrt{D_\alpha}}f_{-1+\alpha/2,\alpha/2}\left(t;\frac{|x-x_0|}{\sqrt{D_\alpha}}\right).
\end{equation}
We mention that $f_{-1+\alpha/2,\alpha/2}$ is called the Mainardi function \cite{pagnini}.

We assume that at the initial moment the subdiffusion parameters are $\alpha$ and $D_\alpha$, and in the long time limit they are $\beta$ and $D_\beta$, $\alpha\neq\beta$. Then, 
\begin{eqnarray}\label{eqV1}
\sigma^2(t)=\left\{
\begin{array}{c}
\frac{2D_\alpha}{\Gamma(1+\alpha)}t^\alpha,\;t\rightarrow 0, \\
   \\
\frac{2D_\beta}{\Gamma(1+\beta)}t^\beta,\;t\rightarrow \infty .
\end{array}
\right.
\end{eqnarray}
Eq. (\ref{eqV1}) coincides with Eq. (\ref{eqII12}) if
\begin{eqnarray}\label{eqV2}
g(t)=\left\{
\begin{array}{c}
t,\;t\rightarrow 0,\\
   \\
At^{\beta/\alpha},\;t\rightarrow \infty,
\end{array}
\right.
\end{eqnarray}
where
\begin{equation}\label{eqV3}
A=\left(\frac{D_\beta \Gamma(1+\alpha)}{D_\alpha \Gamma (1+\beta)}\right)^{\frac{1}{\alpha}}.
\end{equation}
Guided by Eq. (\ref{eqV2}) we propose
\begin{equation}\label{eqV4}
g(t)=a(t)t+(1-a(t))At^{\beta/\alpha},
\end{equation}
where a non--negative function $a$ fulfils the conditions $a(0)=1$, $a(\infty)=0$, and $a$ generates an increasing function $g$ in the time domain. Since $g(t)\rightarrow At^{\beta/\alpha}$ when $t\rightarrow \infty$, Eq. (\ref{eqV4}) provides the additional condition
\begin{equation}\label{eqV5}
t\rightarrow\infty,\;a(t)t \rightarrow 0.
\end{equation}
The function $a$ can be assumed as 
\begin{equation}\label{eqV6}
a(t)=\frac{1}{1+\xi(t)},
\end{equation}
where $\xi$ fulfils the conditions $\xi(0)=0$ and $\xi(\infty)=\infty$. In the following we consider Eq. (\ref{eqV6}) with the power function $\xi\textsl{}(t)=B t^\nu$, where $B$ is a parameter measured in the units of $1/{\rm s}^{1/\nu}$. The condition (\ref{eqV5}) is met for any $\alpha$ and $\beta$, $\alpha,\beta\in(0,1)$, when $\nu>1$. Then, the function $g$ is
\begin{equation}\label{eqV7}
g(t)=\frac{t+ABt^{\frac{\beta}{\alpha}+\nu}}{1+Bt^\nu},
\end{equation}
where $\nu>1$. In this case the Green's function reads 
\begin{eqnarray}\label{eqV8}
P(x,t|x_0)=\frac{1}{2\sqrt{D_\alpha}}\\
\times f_{-1+\alpha/2,\alpha/2}\left(\frac{t+ABt^{\frac{\beta}{\alpha}+\nu}}{1+Bt^\nu};\frac{|x-x_0|}{\sqrt{D_\alpha}}\right),\nonumber
\end{eqnarray}
with $A$ given by Eq. (\ref{eqV3}).

\begin{figure}[htb]
\centering{%
\includegraphics[scale=0.35]{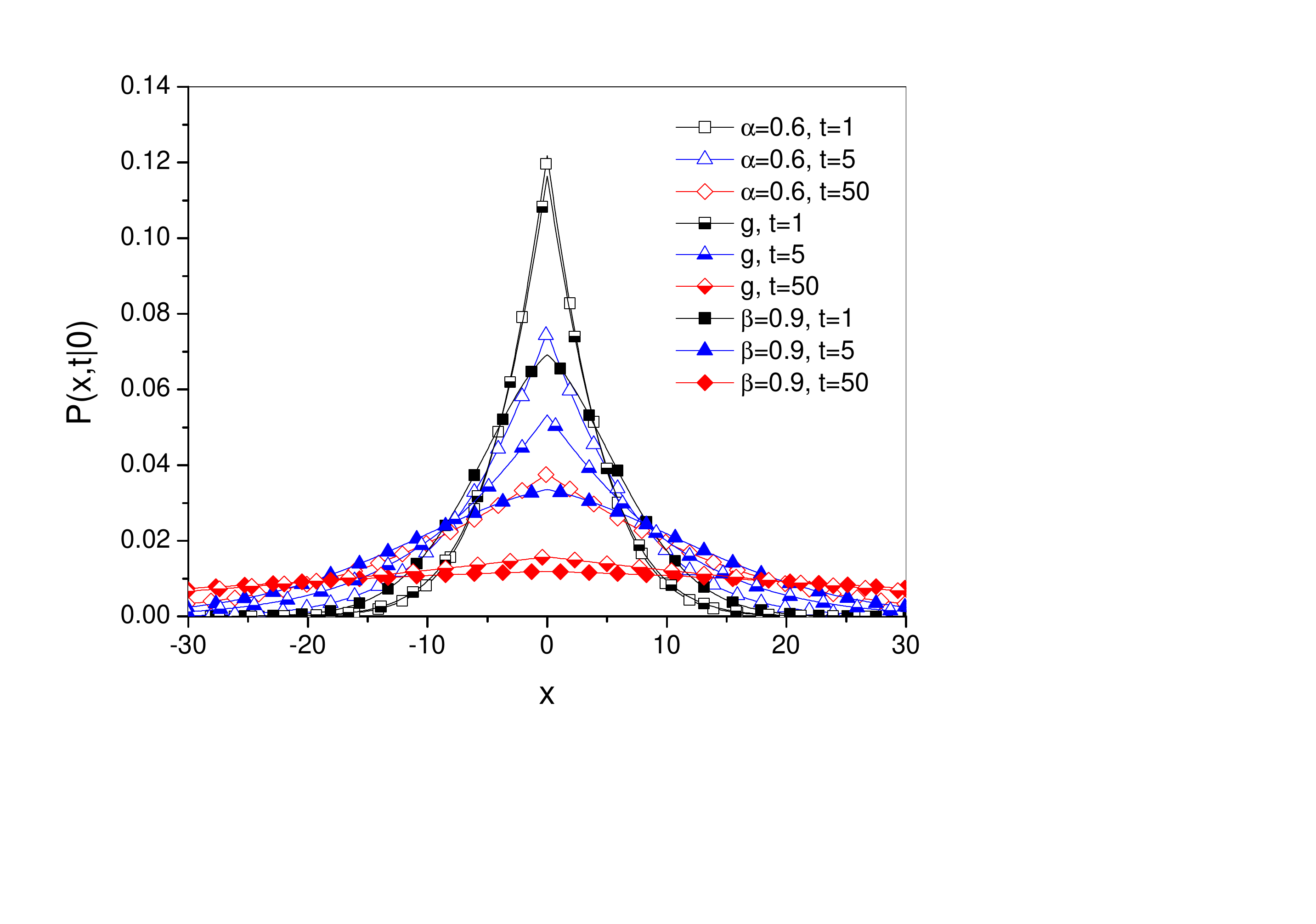}}
\caption{The Green's functions of $g$--subdiffusion equation Eq. (\ref{eqV8}) that describes the transition $(0.6,10)\rightarrow(0.9,20)$ (half full symbols) for $\nu=1.2$. The Green's functions of ``ordinary'' subdiffusion equation Eq. (\ref{eqII13}) are calculated for $\alpha=0.6$ and $D_\alpha=10$ (empty symbols) and for $\beta=0.9$ and $D_\beta=20$ (full symbols). Time values are given in the legend, all quantities are given in arbitrarily chosen units.}
\label{Fig1}
\end{figure}

\begin{figure}[htb]
\centering{%
\includegraphics[scale=0.35]{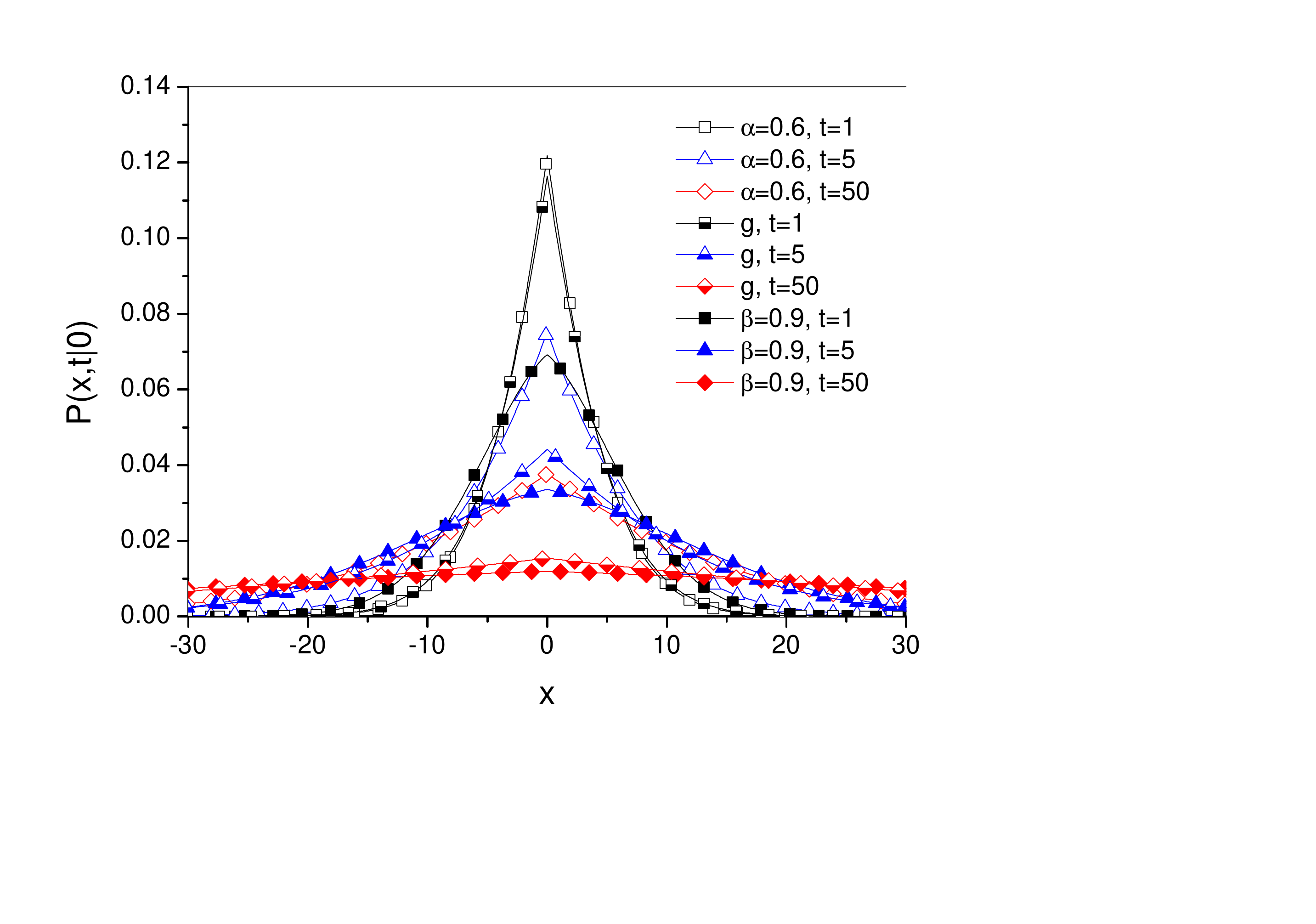}}
\caption{The description is similar as that for Fig. \ref{Fig1} but for $\nu=3.0$.}
\label{Fig2}
\end{figure}

\begin{figure}[htb]
\centering{%
\includegraphics[scale=0.35]{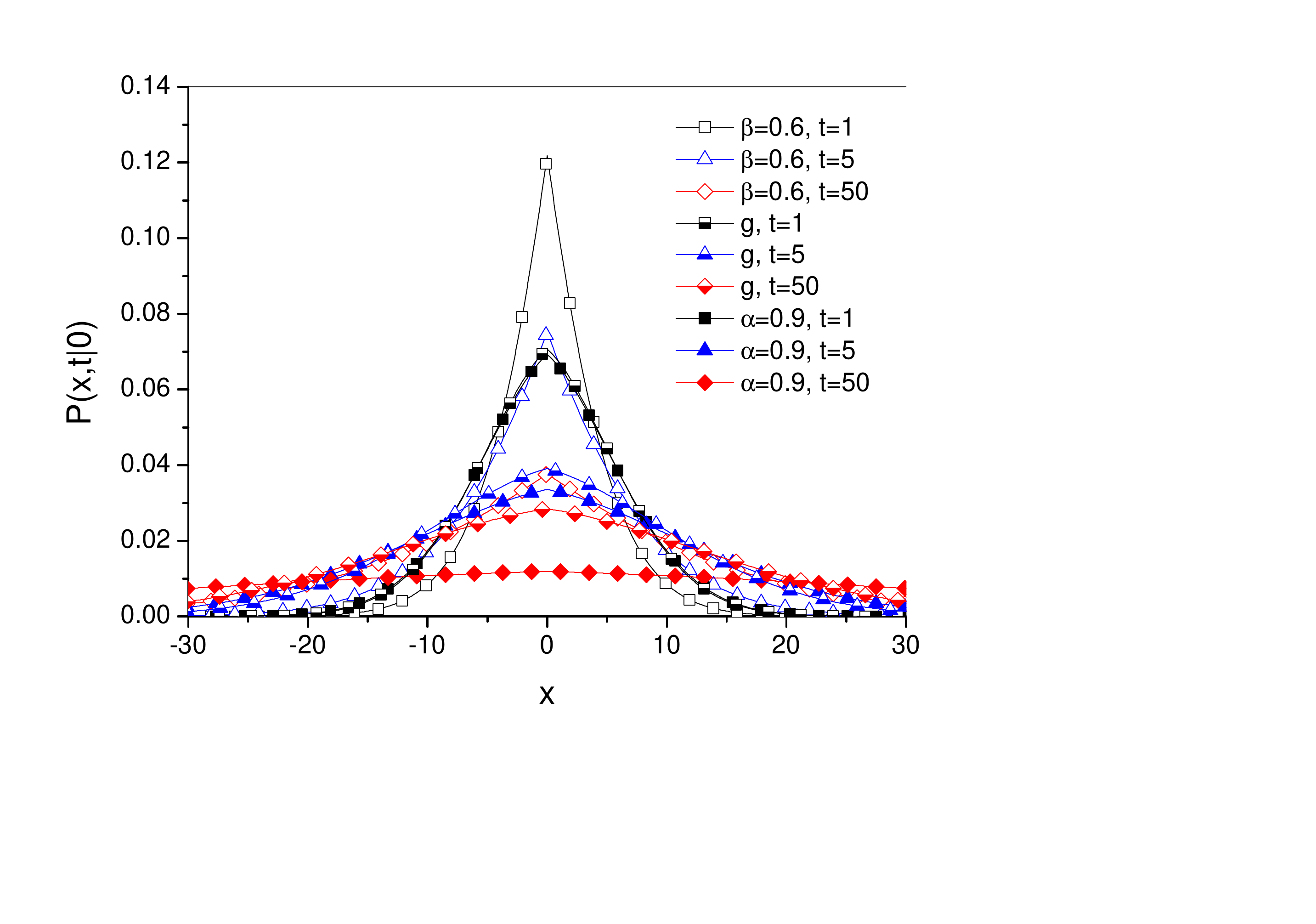}}
\caption{The Green's functions for the process $(0.9,20)\rightarrow(0.6,10)$. The description is similar to that of Fig. \ref{Fig1} for $\nu=1.2$.}
\label{Fig3}
\end{figure}

\begin{figure}[htb]
\centering{%
\includegraphics[scale=0.35]{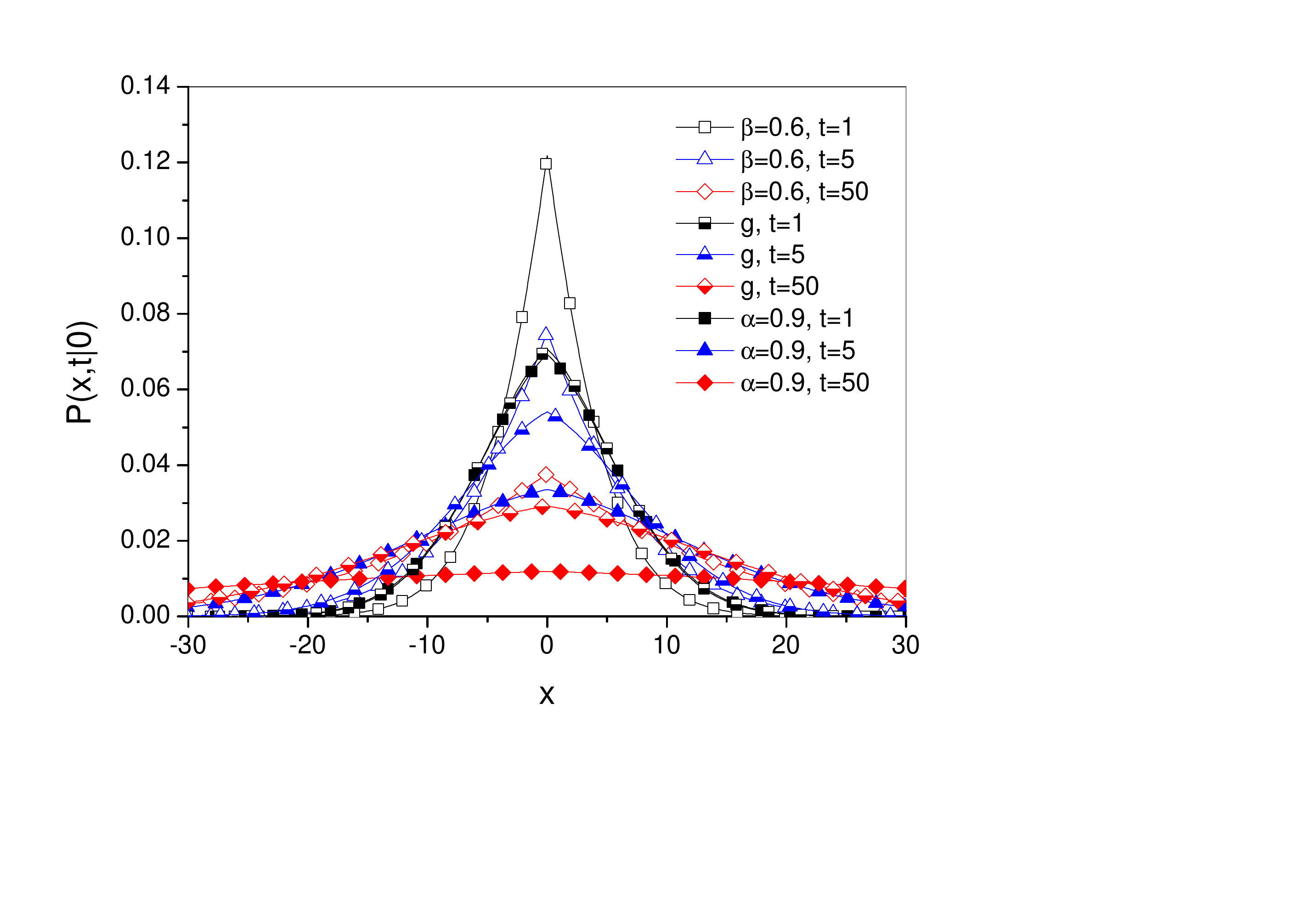}}
\caption{The description is similar as that for Fig. \ref{Fig3} but for $\nu=3.0$.}
\label{Fig4}
\end{figure}

The plots of Green's functions Eq. (\ref{eqV8}) describing the process $(\alpha,D_\alpha) \rightarrow(\beta,D_\beta)$ are compared with the Green's functions for ``ordinary'' subdiffusion with parameters $(\alpha,D_\alpha)$ and $(\beta,D_\beta)$ in Figs. \ref{Fig1}--\ref{Fig4}. We consider accelerated subdiffusion $(0.6,10) \rightarrow(0.9,20)$ and delayed subdiffusion $(0.9,20) \rightarrow(0.6,10)$, both for $B=0.1$ and $x_0=0$, all quantities are given in arbitrarily chosen units. The plots show that for larger $\nu$ $g$--subdiffusion goes to the final process faster. The convergence to the final process seems to be faster for the $(0.6,10)\rightarrow(0.9,20)$ process than for the $(0.9,20)\rightarrow(0.6,10)$ one.

In general, the $g$--subdiffusion equation can be applied to describe subdiffusion for which the MSD time evolution is other than Eq. (\ref{eqI3}).
Let us assume that 
\begin{equation}\label{eqVII1}
\sigma^2(t)=\eta(t), 
\end{equation}
where $\eta$ fulfils the conditions $\eta(0)=0$, $\eta(\infty)=\infty$, and $\eta'(t)>0$ for $t>0$. Comparing Eq. (\ref{eqVII1}) with Eq. (\ref{eqII12}) we find that the $g$--subdiffusion equation Eq. (\ref{eqII1}) with $g(t)=[\Gamma(1+\alpha)\eta(t)/2D_\alpha]^{1/\alpha}$, where $0<\alpha<1$, describes the process which generates Eq. (\ref{eqVII1}).   
As a particular case, subdiffusion with time-varying subdiffusion parameter may be defined by the following relation which is a simple generalization of Eq. (\ref{eqI3}) \cite{sun2010}
\begin{equation}\label{eqVII2}
\sigma^2(t)=\Lambda t^{\tilde{\alpha}(t)},
\end{equation}
$0<\tilde{\alpha}(t)<1$ for $t>0$. It may seem that such a process can be described by the "ordinary" subdiffusion equation with the time-varying order of the fractional derivative \cite{sun2009}
\begin{equation}\label{eqVII3}
\frac{^C \partial^{\tilde{\alpha}(t)} C(x,t)}{\partial t^{\tilde{\alpha}(t)}}=D_\alpha\frac{\partial^2 C(x,t)}{\partial x^2}.
\end{equation}
However, Eq. (\ref{eqVII3}) is difficult to solve, in practice it can be solved numerically \cite{sun2012}. The process which generated Eq. (\ref{eqVII2}) can be described by the $g$--subdiffusion equation with $g(t)=[\Lambda\Gamma(1+\alpha)/2D_\alpha]^{1/\alpha}t^{\tilde{\alpha}(t)/\alpha}$.

The aim of this paper has been to present the $g$--subdiffusion model and its application to describe transient subdiffusion from subdiffusion with parameters $\alpha$ and $D_\alpha$ to subdiffusion with parameters $\beta$ and $D_\beta$. In "intermediate" times the subdiffusive parameters, defined by Eq. (\ref{eqI3}), can remain unknown. We have considered a special case of the function $g$, Eq. (\ref{eqV7}), which describes accelerating subdiffusion when $\alpha<\beta$ and slowing subdiffusion when $\alpha>\beta$. The model uses the $g$--subdiffusion equation with Caputo fractional time derivative with respect to another function $g$.

\end{document}